\RequirePackage{ifpdf}
\ifpdf 
\documentclass[pdftex]{sigma}
\else
\documentclass{sigma}
\fi

\newcommand{\pa}{\partial}

\begin{document}

\allowdisplaybreaks

\renewcommand{\PaperNumber}{025}

\FirstPageHeading

\renewcommand{\thefootnote}{$\star$}

\ShortArticleName{Quantum Super-Integrable Systems as Exactly Solvable Models}

\ArticleName{Quantum Super-Integrable Systems\\ as Exactly Solvable
Models\footnote{This paper
is a contribution to the Vadim Kuznetsov Memorial Issue
``Integrable Systems and Related Topics''. The full collection is
available at
\href{http://www.emis.de/journals/SIGMA/kuznetsov.html}{http://www.emis.de/journals/SIGMA/kuznetsov.html}}}

\Author{Allan P. FORDY}

\AuthorNameForHeading{A.P. Fordy}

\Address{Department of Applied Mathematics, University of Leeds, Leeds LS2 9JT, UK}
\Email{\href{mailto:allan@maths.leeds.ac.uk}{allan@maths.leeds.ac.uk}}
\URLaddress{\url{http://www.maths.leeds.ac.uk/cnls/research/fordy/fordy.html}}

\ArticleDates{Received November 14, 2006, in f\/inal form February
05, 2007; Published online February 14, 2007}

\Abstract{We consider some examples of quantum super-integrable systems and the
associated nonlinear extensions of Lie algebras.  The intimate relationship
between super-integrability and exact solvability is illustrated.
Eigenfunctions are constructed through the action of the commuting operators.
Finite dimensional representations of the quadratic algebras are thus
constructed in a way analogous to that of the highest weight representations of
Lie algebras.}


\Classification{35Q40; 70H06}

\begin{flushright}
\it Dedicated to the memory of Vadim Kuznetsov
\end{flushright}

\renewcommand{\thefootnote}{\arabic{footnote}}
\setcounter{footnote}{0}

\section{Introduction}

In classical mechanics, there is a well def\/ined meaning to the term `complete
integrability'.  If, in $n$ degrees of freedom, we have $n$ independent
functions {\it in involution} (mutually Poisson commuting), then the system can
be integrated, `up to quadrature'.  This is known as {\it Liouville's
Theorem}.  It is customary to consider one of these functions as `the
Hamiltonian' $H$, with the others being its {\it first integrals} and this
Hamiltonian is said to be {\it completely integrable in the Liouville sense}.
Whilst $n$ is the maximal number of independent functions which can be {\it in
involution}, it is possible to have further integrals of the Hamiltonian $H$,
which necessarily generate a non-Abelian algebra of integrals of $H$.  The
maximal number of additional {\it independent} integrals is $n-1$, since the
`level surface' of $2n-1$ integrals (meaning the intersection of individual
level surfaces) is just the (unparameterised) integral curve.  Well known
elementary examples are the isotropic harmonic oscillator, the Kepler system
and the Calogero--Moser system.  The quadratures of complete integrability are
often achieved through the separation of variables of the Hamilton--Jacobi
equation.  The solution of a maximally super-integrable system can also be
calculated purely algebraically (albeit implicitly), requiring just the
solution of the equations $I_k=c_k$, $k=1,\ldots,2n-1$.

In $n$-dimensions, Quantum integrable systems are def\/ined analogously by
requiring the exis\-tence of $n$ mutually commuting dif\/ferential operators
(usually with some requirements, such as self-adjointness).  One of these will
be of Schr\"odinger type, being the quantum version of the Hamiltonian $H$.
However, we don't have a theorem analogous to Liouville's and we don't have
anything resembling `reduction to quadrature'. There {\it is} a notion of
separation of variables (this time in the linear PDE sense), requiring a
solution to be expressible as the product of $n$ functions of a single
variable.  This reduces the problem of solving one Schr\"odinger equation in
$n$-dimensions to that of solving $n$ Schr\"odinger equations in 1-dimension.
This achieves {\it less} than quadratures.

What we would like to achieve in quantum mechanics is to build the spectrum and
corresponding eigenfunctions for Schr\"odinger's equation with some given
boundary conditions.  This class of Schr\"odinger equation is called {\it
exactly solvable}.  Even in 1-dimension exactly solvable Schr\"odinger
equations are rare, so unlikely to be the end-product of applying the
separation of variables method.  One particular class of 1-dimension exactly
solvable Schr\"odinger equations was presented by Infeld and Hull \cite{51-1},
related to the {\it factorisation of operators}.  This can be related to
Darboux transformations, which can then be generalised to higher dimensions
(where factorisation is not possible).  Darboux transformations can then be
used to build exactly solvable Schr\"odinger equations in higher dimensions. In
\cite{f06-1}, Darboux transformations have been shown to be related to the
symmetries of the Laplacian.

In the quantum case we also have a notion of {\it super-integrability}.  By
analogy with the classical case, super-integrability again involves having
`extra' commuting operators.  Once again, we cannot have more than $n$ such
operators {\it in involution}, so the algebra is necessarily non-Abelian.
Furthermore, if the operators are second order, their commutators are third
order, with further commutators being fourth order and so on, so cannot be
expected to generate a {\it finite dimensional Lie algebra}.  However, they
{\it can} generate a f\/inite dimensional algebra with {\it nonlinear commutation
relations} (see \cite{01-5,99-11,96-5}, for example).  There are many papers on
super-integrable systems (both classical and quantum).  Particularly relevant
to the present paper is a series of papers by Kalnins et al. (see
\cite{01-7, 99-11} and references therein), in which a classif\/ication is given
of super-integrable systems in 2-dimensional f\/lat and constant curvature
spaces and in which a pair of quadratic f\/irst integrals, together with their
quadratic commutation relations are derived.  As is well known,
super-integrable systems are separable in more than one coordinate system.  In
most of these papers the main emphasis is on the separation of variables, f\/irst
by constructing and classifying coordinate systems and corresponding
`separable potentials' and then using this (in the quantum case) to build
eigenfunctions.

In the present paper, there is no discussion of separation of variables and
`dif\/ferential equations techniques' are kept to a minimum.  Our main emphasis
is on the {\it direct relationship} between {\it super-integrability} and {\it
exactly solvability}.  We use the commuting operators in a {\it direct and
explicit manner} to build the eigenfunctions for our operators.  This
construction is the analogue (in the context of our quadratic algebras) to that
of the {\it highest weight representation} of simple Lie algebras (see the
Appendix). In particular, the polynomial eigenfunctions of the Krall--Shef\/fer
operator of Example \ref{superhe} can be {\it explicitly constructed} in this
way.  The issue of the relationship between super-integrability and exact
solvability is not new and is discussed in \cite{99-11,06-1,01-8}.  In
\cite{01-8} it is conjectured that exactly solvability should hold quite widely
for super-integrable systems.  Whilst operator algebras are discussed in these
papers they are not used to construct eigenfunctions.

\looseness=-1
In this paper we consider some particular super-integrable systems, associated
with a particular Laplace--Beltrami operator in 2-dimensions.  One of the
examples is in the Krall--Shef\/fer classif\/ication of `admissible operators'.
The relationship between Krall--Shef\/fer operators and super-integrability was
shown by Harnad, et al in \cite{01-1,03-4}. In these papers they reworked the
Krall--Shef\/fer classif\/ication and showed that in each case the leading order
terms in the ope\-ra\-tor correspond to the Laplace--Beltrami operator of either a
zero or constant curvature, 2-di\-mensional space.  They also show that each of
the $9$ cases is super-integrable by presenting pairs of commuting, second
order operators. Example \ref{superhe} of the current paper is {\it exactly}
Case II of \cite{01-1,03-4}.  In \cite{03-4} it is pointed out that this case
had been previously introduced and studied in \cite{83-9,99-11}.

In the next section, we give some basic formulae related to our particular
metric and its symmetries.  We then present $3$ super-integrable examples, with
their operator algebras and f\/inite representation spaces (families of
eigenfunctions).  We then give some concluding remarks.

\section{A metric and its symmetries}  \label{sec-metric}

For an $n$-dimensional (pseudo-)Riemannian space, with local coordinates
$x^1,\dots,x^n$ and metric~$g_{ij}$, the Laplace--Beltrami operator is def\/ined
by $L_b f = g^{ij} \nabla_i\nabla_j f$, which has explicit form  %
\begin{gather}
\label{lbf}   %
L_b f = \sum_{i,j=1}^n \frac{1}{\sqrt{g}}\frac{\pa}{\pa x^j}
\left(\sqrt{g} g^{ij}\frac{\pa f}{\pa x^i}\right),
\end{gather}  %
where $g$ is the determinant of the matrix $g_{ij}$.  The coef\/f\/icients of
leading order terms in the Lap\-lace--Beltrami operator are the coef\/f\/icients of
the {\it inverse metric} $g^{ij}$.  For a metric with isometries, the
inf\/initesimal generators (Killing vectors) are just f\/irst order dif\/ferential
operators which commute with the Laplace--Beltrami operator (\ref{lbf}).  When
the space is either f\/lat or constant curvature, it possesses the maximal group
of isometries, which is of dimension $n(n+1)/2$.  In this case, $L_b$
is actually the second order {\it Casimir} function of the symmetry algebra
(see \cite{74-7}).

In this paper, we consider one particular constant curvature metric in
2-dimensions, with inverse  %
\begin{gather}
\label{metric}  %
g^{ij} = \left(\begin{matrix}
x^2 & x y
\\
x y & y^2-y
\end{matrix}\right).
\end{gather}%
The choice of coordinates here is motivated by the relationship to
Krall--Shef\/fer operators \cite{67-2}, but otherwise quite arbitrary.

A convenient basis of Killing vectors is   %
\begin{gather}
\label{sl2-1} %
{\bf H} = 4 x \pa_x,\qquad
{\bf E} = 2 \sqrt{x y} \pa_y,\qquad
{\bf F} = 4 \sqrt{x y} \pa_x + 2 (y-1) \sqrt{\frac{y}{x}}\pa_y,
\end{gather}%
satisfying the standard commutation relations of $sl(2,\mathbb{C})$:  %
\begin{gather}
\label{sl2comm} %
[{\bf H},{\bf E}] = 2 {\bf E}, \qquad
[{\bf H},{\bf F}] = -2 {\bf F}, \qquad
[{\bf E},{\bf F}] =  {\bf H}.
\end{gather}%
The Laplace--Beltrami operator for the metric (\ref{metric}) is proportional to
the quadratic Casimir operator:    %
\begin{gather}
\label{cas}   %
L_b = \frac{1}{16} ( {\bf H}^2 + 2 {\bf E}{\bf F} + 2 {\bf F}{\bf E}) =
x^2 \pa_x^2 + 2 x y \pa_x \pa_y + (y^2-y) \pa_y^2 + \frac{3}{2} x \pa_x
+ \frac{1}{2} (3 y-1) \pa_y.  %
\end{gather}  %

\begin{remark}  The operator $L_b$ is invariant under an {\bf involution},  %
\begin{gather}
\label{involx}      %
\bar x = \frac{(y-1)^2}{x},\qquad \bar y = y.
\end{gather}      %
which, in fact, gives a concrete realisation of the Lie algebra automorphism
\begin{gather*}
{\bf E} \leftrightarrow {\bf F}, \qquad {\bf H} \rightarrow -{\bf H}.
\end{gather*}
This will be useful in some of the calculations below.   %
\end{remark}

Second order operators, commuting with $L_b$, are just symmetric quadratic
forms of Killing vectors.  Suppose $\bf K$ is such an operator.  Then we may
seek functions $u$ and $v$, such that
\begin{gather*}
[L_b+u,{\bf K}+v]=0,
\end{gather*}
which constitutes a coupled system of partial dif\/ferential equations for $u$
and $v$.  The solution depends upon a pair of arbitrary functions, each of one
variable, typical of separable systems.  Requiring that there exist two of
these second order commuting operators, $I_j={\bf K}_j+v_j$, strongly
constrains these (formerly) arbitrary functions, which reduce to rational
functions depending upon only a f\/inite number of parameters.  The coef\/f\/icients
of the second order derivatives in such an operator $\bf K$ def\/ine a
contravariant, rank-two Killing tensor.  For brevity, we will refer to these
operators as Killing tensors in what follows.

The examples discussed below were presented in \cite{f06-1}, in the context of
Darboux transformations, which will not be discussed in this paper.  Instead we
discuss how to use the commuting second order operators to build eigenfunctions
of $L=L_b+u$ in a way analogous to the highest weight representation of a Lie
algebra.

For some calculations, such as those of \cite{f06-1}, it is convenient to use
separation coordinates.  In our case we can choose $2$ (or even $3$) such
coordinate systems, but the current calculations are as easily done in terms of
the $x-y$ coordinates.

\section{The system with Killing tensors ${\bf H}^2$ and ${\bf E}^2$}

Requiring {\it both} $I_1={\bf H}^2+v_1$ and $I_2={\bf E}^2+v_2$ to commute
with $L$ leads to the specif\/ic forms of $u(x,y)$, $v_i(x,y)$ in the operators
below
\begin{gather}
L = L_b + \frac{c_0}{x} + \frac{c_1}{y} + c_2\frac{y-1}{x^2},\nonumber
\\
I_1 = {\bf E}^2 - 4 \left(c_1\frac{x}{y} + c_2 \frac{y}{x}\right),\nonumber
\\
I_2 = {\bf H}^2 - 16 \left( c_0 \left(\frac{y-1}{x}\right) +
c_2 \left(\frac{y-1}{x}\right)^2 \right).
\label{ueh}
\end{gather}
In order to build eigenfunctions of $L$ it is convenient to change gauge:
\begin{gather}
\label{gauge}
L \mapsto \tilde L = G^{-1} LG \qquad\mbox{with}\qquad
G = \exp\left(\frac{\alpha}{2}\left(\frac{y-1}{x}\right)\right)
x^{(\beta+\gamma-1)/2} y^{-(2\gamma+1)/4},
\end{gather}
which simultaneously reduces the $3$ potential terms to constants.  The
specif\/ic coef\/f\/icients in $G$ were chosen to simplify the f\/inal expressions for
the operators.  Removing the additive constants and adjusting overall
multiplicative factors (and dropping the `tilde') we arrive at the following
operators: 
\begin{gather}
 L = x^2 \pa_x^2 + 2 x y \pa_x\pa_y + \big(y^2-y\big) \pa_y^2
 + (\beta x + \alpha)\pa_x + (\beta y + \gamma)\pa_y,\nonumber
 \\
I_1 = x y \pa_y^2 + (\alpha y - \gamma x) \pa_y,\nonumber
\\
I_2 = x^2 \pa_x^2 +((\beta+\gamma) x+\alpha (1-y)) \pa_x,
\label{superhe}
\end{gather}%
where $c_0=\alpha (2-\beta-\gamma)/2$, $c_1=(2\gamma+1)(2\gamma+3)/16$,
$c_2=\alpha^2/4$.  This is an example of {\it admissible operator}, as def\/ined
by Krall and Shef\/fer \cite{67-2}. The connection between Krall--Shef\/fer
operators and super-integrability was f\/irst pointed out in \cite{01-1,03-4}.
Indeed, Example~\ref{superhe} above is {\it exactly} Case~II of
\cite{01-1,03-4}. As pointed out in \cite{03-4}, this case was previously
introduced and studied in~\cite{83-9,99-11}.

The operator $I_3=[I_1,I_2]$ is third order, so cannot be written as a
polynomial (in particular, {\it linear}, Lie algebraic) expression in $L,\,
I_1,\, I_2$, but does satisfy the polynomial equation:
\begin{gather*}
I_3^2 =\frac{4}{3}(I_1^2I_2+I_1I_2I_1+I_2I_1^2-2 I_1^2) +
\alpha(\beta-\gamma-2)\left(I_1I_2+I_2I_1-\frac{4}{3}I_1\right)
\\  \phantom{I_3^2 =}{}
{}+ (\beta+\gamma-2) ((\beta+\gamma)I_1^2-2\alpha I_1L)
+\alpha^2 (I_2-L)^2+\frac{2}{3}\alpha^2(I_2-(3\gamma+1)L).
\end{gather*}
These $4$ operators satisfy the commutation relations:
\begin{gather}
[L,I_1]=[L,I_2]=[L,I_3]=0,\qquad [I_1,I_2]=I_3, \nonumber
\\
[I_1,I_3]=2 I_1^2+\alpha (\beta-\gamma-2) I_1 +\alpha^2 (I_2 -L),\nonumber
\\
[I_2,I_3]= -2(I_1I_2+I_2I_1)+(\beta+\gamma-2)(\alpha L-(\beta+\gamma)I_1)-\alpha (\beta-\gamma-2)I_2 .
\label{he-comm}
\end{gather}

By def\/inition (see \cite{67-2}), an {\it admissible operator} possesses a
sequence of eigenvalues $\lambda_n$ (with $\lambda_n\neq\lambda_m$ for $m\neq
n$)\footnote{The condition (3.8) of \cite{67-2}, which ensures this, is that
$\beta$ should not be a negative integer.}, such that, for each $n\ge 0$, there
exist $n+1$ linearly independent polynomial eigenfunctions of degree $n$, with
eigenvalue $\lambda_n$.  The {\it degeneracy} (for $n>0$) of $\lambda_n$ stems
from the super-integrability in a very explicit way.  We show below how to use
the operators $I_1$ and $I_2$ to build sequences of polynomial eigenfunctions
of $L$, using a construction which is analogous to that of the `highest weight
representation' of a simple Lie algebra.

We start with a polynomial eigenfunction
\begin{gather*}
P_{n,0} = x^n +\sum_{i=1}^n \rho_i x^{n-i},
\end{gather*}
for constants $\rho_i$, which is independent of $y$ and therefore in the Kernel
of $I_1$.  This satisf\/ies the 1-dimensional eigenvalue problem
\begin{gather*}
L_x P_{n,0} \equiv (x^2 \pa_x^2 + (\beta x + \alpha)\pa_x)P_{n,0} = \lambda_n
P_{n,0}, \qquad \lambda_n= n(n+\beta-1).
\end{gather*}
This value of $\lambda_n$ is easily determined by looking at the coef\/f\/icient of
$x^n$. It is easy to show that
\begin{gather*}
P_{1,0}=x+\frac{\alpha}{\beta},\qquad P_{2,0}=x^2+\frac{2\alpha
x}{\beta+2}+\frac{\alpha^2}{(\beta+2)(\beta+1)},
\end{gather*}
and that this sequence satisf\/ies the 3-point recursion relation
\begin{gather*}
{\bf r}_x:\quad P_{n+1,0}= (x+A_n) P_{n,0} + B_n P_{n-1,0},
\end{gather*}
with
\begin{gather*}
   A_n=\frac{\alpha (\beta-2)}{(\beta+2n)(\beta+2n-2)},\qquad
   B_n=\frac{\alpha^2 n (n+\beta-2)}{(\beta+2n-1)(\beta+2n-2)^2(\beta+2n-3)}.
\end{gather*}

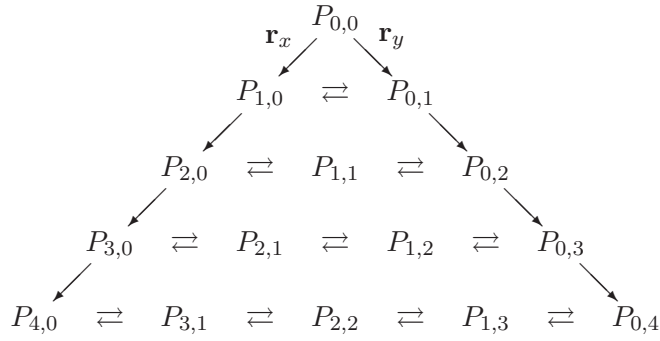
\begin{figure}[hbt]
\begin{center}
\unitlength=0.5mm
\begin{picture}(160,80)
\put(80,80){\makebox(0,0){$P_{0,0}$}}
\put(60,60){\makebox(0,0){$P_{1,0}$}}
\put(100,60){\makebox(0,0){$P_{0,1}$}}
\put(40,40){\makebox(0,0){$P_{2,0}$}}
\put(80,40){\makebox(0,0){{$P_{1,1}$}}}
\put(120,40){\makebox(0,0){$P_{0,2}$}}
\put(20,20){\makebox(0,0){$P_{3,0}$}}
\put(60,20){\makebox(0,0){$P_{2,1}$}}
\put(100,20){\makebox(0,0){$P_{1,2}$}}
\put(140,20){\makebox(0,0){$P_{0,3}$}}
\put(0,0){\makebox(0,0){$P_{4,0}$}}
\put(40,0){\makebox(0,0){{$P_{3,1}$}}}
\put(80,0){\makebox(0,0){$P_{2,2}$}}
\put(120,0){\makebox(0,0){{$P_{1,3}$}}}
\put(160,0){\makebox(0,0){$P_{0,4}$}}
\put(75,75){\vector(-1,-1){10}}
\put(85,75){\vector(1,-1){10}}
\put(55,55){\vector(-1,-1){10}}  \put(105,55){\vector(1,-1){10}}
\put(35,35){\vector(-1,-1){10}}  \put(125,35){\vector(1,-1){10}}
\put(15,15){\vector(-1,-1){10}}  \put(145,15){\vector(1,-1){10}}
\put(80,60){\makebox(0,0){$\rightleftarrows$}}
\put(60,40){\makebox(0,0){$\rightleftarrows$}}
\put(100,40){\makebox(0,0){$\rightleftarrows$}}
\put(40,20){\makebox(0,0){$\rightleftarrows$}}
\put(80,20){\makebox(0,0){$\rightleftarrows$}}
\put(120,20){\makebox(0,0){$\rightleftarrows$}}
\put(20,0){\makebox(0,0){$\rightleftarrows$}}
\put(60,0){\makebox(0,0){$\rightleftarrows$}}
\put(100,0){\makebox(0,0){$\rightleftarrows$}}
\put(140,0){\makebox(0,0){$\rightleftarrows$}}
\put(65,75){\makebox(0,0){${\bf r}_x$}}
\put(95,75){\makebox(0,0){${\bf r}_y$}}
\end{picture}
\end{center}
\caption{The triangular lattice of polynomials $P_{n-k,k}$, $k=0,\dots, n$
with $P_{0,0}=1$, for system (\ref{superhe}).  Horizontal arrows denote the
action of $I_1$ (left) and $I_2$ (right).} \label{trilatt}
\end{figure}

We now use $I_2$ (for $\alpha \neq 0$) to build a sequence of ({\it monic})
eigenfunctions
\begin{gather*}
P_{n-k,k} = x^{n-k}y^k + \;\mbox{lower order terms},\qquad k=1,\ldots,n.
\end{gather*}
The case of $\alpha=0$ is the reduction $c_0=c_2=0$, which has a dif\/ferent
algebra and is dealt with later (see the system (\ref{uef})).  Since $I_2$
commutes with $L$, all such eigenfunctions will have the same eigenvalue
$\lambda_n$, which means (important for $P_{0,n}$, below) that the degree of
each polynomial must be $n$. The polynomials $P_{n-k,k}$ are def\/ined
recursively by
\begin{gather}
\label{i2p}   %
I_2 P_{n-k,k}\!=\!(n\!-\!k)(n\!-\!k\!-\!1\!+\!\beta\!+\!\gamma) P_{n-k,k}\!-\!\alpha (n\!-\!k)P_{n-k-1,k+1}, \qquad
k=0,\ldots, n-1,
\end{gather}   %
as can be seen by inspecting the form of $I_2$ and by inspecting the
coef\/f\/icients of $x^{n-k}y^k$ and $x^{n-k-1}y^{k+1}$.  It can be seen that $I_2
P_{0,n}=0$, so the polynomial $P_{0,n}$ must be independent of
$x$.\footnote{Since the leading order term is killed, any non-zero result would
be a polynomial of lower degree, which is not possible. This can also be seen
by setting $k=n$ in (\ref{i2p}).}  The polynomials $P_{0,n}(y)$ also satisfy a
3-point recursion relation \cite{f05-1}
\begin{gather*}
{\bf r}_y : \quad P_{0,n+1}= (y+\tilde A_n) P_{0,n} + \tilde B_n P_{0,n-1},
\end{gather*}
with
\begin{gather*}
\tilde A_n = \frac{(n+1)(\gamma-n)}{\beta+2 n}- \frac{n(\gamma-n+1)}{\beta+2 n-2},
\\
\tilde B_n = -\frac{n (\gamma-n+1)}{2 (\beta+2 n-2)}
\left((n+1)\frac{\gamma-n}{\beta+2 n-1} - 2n\frac{\gamma-n+1}{\beta+2 n-2} +
(n-1)\frac{\gamma-n+2}{\beta+2 n-3}\right).
\end{gather*}
The space of polynomials spanned by this basis is also invariant under the
action of $I_1$:
\begin{gather*}
I_1 P_{n-k,k} = k\alpha P_{n-k,k} +k (k-\gamma-1)P_{n-k+1,k-1},
   \qquad k=1,\dots, n.
\end{gather*}
This array of polynomials, together with the action of $I_1$, $I_2$ and the
3-point recursion relations, are depicted in Fig.~\ref{trilatt}.

For each $n\geq 1$, the sequence of polynomials $P_{n-k,k}$, $k=0,\dots,n$,
def\/ines an $n+1$ dimensional representation of the algebra (\ref{he-comm}).
These can be explicitly calculated.  For instance, apart from~$P_{1,0}$ and
$P_{2,0}$, given above, we have
\begin{gather*}
 P_{0,1} = y + \frac{\gamma}{\beta},\qquad
    P_{1,1} = x y + \frac{1}{\beta+2}(\gamma x+\alpha y) +
             \frac{\alpha\gamma}{(\beta+2)(\beta+1)},\\
  P_{0,2} = y^2 + 2\, \frac{(\gamma-1) y}{\beta+2} +
               \frac{\gamma(\gamma-1)}{(\beta+2)(\beta+1)} .
\end{gather*}
It is easy to see that, for $n\geq 2$, the $4$ representative matrices generate
a larger Lie algebra (commutators lie outside the linear span), but {\it do}
satisfy the {\it quadratic} relations (\ref{he-comm}).

\section{The system with Killing tensors ${\bf H}^2$ and ${\bf F}^2$}

Requiring {\it both} $I_1={\bf H}^2+v_1$ and $I_2={\bf F}^2+v_2$ to commute
with $L$ leads to the specif\/ic forms of~$u(x,y)$, $v_i(x,y)$ in the operators
below
\begin{gather*}
L = L_b + c_0 \frac{x}{(y-1)^2} + \frac{c_1}{y} +  c_2\, \frac{x^2}{(y-1)^3},\nonumber
\\
I_1 = {\bf F}^2 - 4 \left( c_1 \frac{(y-1)^2}{x y} + c_2 \frac{x y}{(y-1)^2}\right), \nonumber
\\
I_2 = {\bf H}^2 - 16 \left(c_0 \left(\frac{x}{y-1}\right) +
c_2 \left(\frac{x}{y-1}\right)^2\right).
\end{gather*}
These are, in fact, directly transformed from (\ref{ueh}) under the action of
the involution (\ref{involx}).  Either by composing the gauge transformation
(\ref{gauge}) with this involution or by directly transforming (\ref{superhe})
by (\ref{involx}) (and dropping `bars'), we obtain
\begin{gather*}   %
L = x^2 \pa_x^2 + 2 x y \pa_x\pa_y + (y^2-y) \pa_y^2
+ \left(\beta x + \frac{2(\beta+\gamma-1)x}{y-1}-
\frac{\alpha x^2}{(y-1)^2}\right)\pa_x + (\beta y + \gamma) \pa_y,\nonumber
\\
I_1 = 4 x y \pa_x^2 + 4 y(y-1) \pa_x\pa_y + \frac{y (y-1)^2}{x}\pa_y^2
+2\left((1 - \gamma)y+ \frac{\alpha x y}{y-1} +\gamma\right) \pa_x\nonumber
\\ \phantom{I_1 =}
 {}+\left(\alpha y -\frac{\gamma (y-1)^2}{x}\right)\pa_y,\nonumber
\\
I_2 = x^2 \pa_x^2 +\left((2-\beta-\gamma) x-\frac{\alpha x^2}{1-y}\right) \pa_x ,
\end{gather*}%
where $c_0=\alpha (2-\beta-\gamma)/2$, $c_1=(2\gamma+1)(2\gamma+3)/16$,
$c_2=\alpha^2/4$.  These satisfy the same quadratic relations (\ref{he-comm}).
Starting from these operators, it would not be so easy to f\/ind a simple
function in the kernel of $I_1$, and then to build the sequence of
eigenfunctions, but there is no need to do this, since the functions
\begin{gather*}
Q_{n-k,k}(x,y)=P_{n-k,k}\left(\frac{(y-1)^2}{x},y\right)
\end{gather*}
are obtained directly through the involution.  The operator $L$ is no longer
`admissible' in the Krall--Shef\/fer sense, and the eigenfunctions are no longer
polynomial, but the same set of relations hold, leading to the same diagram of
Fig.~\ref{trilatt}.

\section{The system with Killing tensors ${\bf E}^2$ and ${\bf F}^2$}

Requiring {\it both} $I_1={\bf E}^2+v_1$ and $I_2={\bf F}^2+v_2$ to commute
with $L$ leads to the specif\/ic forms of $u(x,y)$, $v_i(x,y)$ in the operators
below   %
\begin{gather}
\label{uef}        %
L = L_b + \frac{c_1}{y},  \qquad I_1 = {\bf E}^2 - 4 c_1\, \frac{x}{y}, \qquad
I_2 = {\bf F}^2 - 4 c_1 \,\frac{(y-1)^2}{x y}.
\end{gather}    %
However, it can be seen that the potential of $L$ is a subcase of (\ref{ueh}),
with $c_0=c_2=0$, so there should also be an integral of the form $I_3={\bf
H}^2+v_3$.  Indeed, it should be {\it exactly} as in the case of (\ref{ueh}).
However, in this reduction, we just have $I_3={\bf H}^2$, so may consider the
algebra generated by $L$, $I_1$, $I_2$ and $\bf H$, which satisfy the
commutation relations
\begin{gather}
\label{ef-comm}    %
[{\bf H},I_1]=4 I_1,\qquad  [{\bf H},I_2]= -4 I_2,\qquad
  [I_1,I_2] = 2 (8c_1-1) {\bf H} - {\bf H}^3 +16 {\bf H}\, L,
\end{gather}   %
as well as $[L,I_1]=[L,I_2]=[L,{\bf H}]=0$.  Under the action of the involution
(\ref{involx}), $L\mapsto L$, $I_1\mapsto I_2$, $I_2\mapsto I_1$, ${\bf
H}\mapsto - {\bf H}$, which preserves the commutation relations
(\ref{ef-comm}).  The Casimir operator of the algebra (\ref{ef-comm}) is
\begin{gather*}
{\cal C} = I_1I_2+I_2I_1+\frac{1}{2} (8c_1-5){\bf H}^2-\frac{1}{8} {\bf H}^4 +4
{\bf H}^2 L.
\end{gather*}
This is not, of course, independent of $L$ and can be written ${\cal C}=32
L^2-16 (4 c_1+1)L+16 c_1(2c_1-1)$.

We mimic the highest weight representation of $sl(2,\mathbb{C})$ and seek
function $\varphi_1(x,y)$, satisfying
\begin{gather*}
{\bf H}\varphi_1=s\varphi_1,\qquad L\varphi_1=\lambda\varphi_1,\qquad
I_1\varphi_1=0.
\end{gather*}
The f\/irst of these gives $\varphi_1(x,y)=x^{s/4} \psi(y)$, so the remaining
equations lead to
\begin{gather*}
16 y^2(y-1)\psi'+8y((s+3)y-1)\psi'+(16c_1+(s(s+2)-16\lambda)y)\psi=0,\\
    2y^2\psi'+y\psi'-2c_1\psi=0,
\end{gather*}
which can be solved for both $\lambda$ and $\psi$ to give
\begin{gather*}
\lambda_{m,s}=\frac{1}{16}(s+m+1)(s+m+3),\qquad
\psi=y^{(m+1)/4},\qquad\mbox{where}\qquad
c_1=\frac{m^2-1}{16}.
\end{gather*}
We now def\/ine
\begin{gather*}
\varphi_k = I_2^{k-1}\varphi_1,\qquad\mbox{satisfying}\qquad
     {\bf H}\varphi_k= (s-4(k-1))\varphi_k.
\end{gather*}
Under the involution, we have $\varphi_1\mapsto \bar \varphi_1$ and
\begin{gather*}
{\bf H}\bar \varphi_1 = -s \bar \varphi_1,\qquad I_2 \bar \varphi_1 = 0.
\end{gather*}
For this to be in our sequence $\varphi_k$, we must have $s-4(k-1)=-s$ for some
$k$.  For some integer $n$, we therefore have
\begin{gather*}
s = 2n \qquad\mbox{and}\qquad \bar \varphi_1 = \varphi_{n+1}, \qquad
           I_2\varphi_{n+1} = 0.
\end{gather*}
In this case, $\varphi_k,\; k=1,\ldots,n+1$, def\/ine a {\it finite dimensional}
representation of the algebra.  For other values of $s$, the representation is
inf\/inite.

The representations obtained in this way are {\it deformations} of
representations of $sl(2,\mathbb{C})$ (given in the Appendix), which correspond
to the reduction $c_1=0$ (which is $m=-1$). In this reduction, the eigenvalue
reduces to the value $\lambda_{-1,2n}=\frac{1}{4}n(n+1)$.  When $c_1=0$, $I_2$
reduces to ${\bf F}^2$, so our sequence $\varphi_1, \varphi_2, \dots$
reduces to $\psi_1^n, \psi_3^n, \dots$, jumping in steps of $2$.  We thus
retrieve only $n+1$ of the $2n+1$ vectors of the highest weight representation.

\begin{example}[The case $n=2$]  {\it   %
Here we have
\begin{gather*}
\varphi_1 = x y^p,\qquad \varphi_2= 4(3y-1+4p(y-1))y^p,\qquad \varphi_3= 8(y-1)^2
(3+16p+16p^2)y^p,
\end{gather*}
where $m=-1+4p$.  When $p=0$, these reduce to $\psi_1^2$, $\psi_3^2$ and
$\psi_5^2$, given in the Appendix.
}
\end{example}

\section{Conclusions}

The main message of this paper is that {\it super-integrability} in a quantum
system leads {\it directly and explicitly} to the construction of
eigenfunctions.  This leads to {\it exact solvability}, which is much stronger
than just {\it complete integrability}.

The examples presented here are fairly simple, but the general approach can be
used for {\it any} super-integrable system.  The examples presented here are
associated with the Laplace--Beltrami operator of a 2-dimensional space of
constant curvature.  The main feature used was the existence of a large number
of symmetries of the underlying space.  Using these it is possible to deform
the Laplace--Beltrami operator, by adding a potential, in such a way that it
possesses (generally) higher order (non-geometric) symmetries in the form of
higher order commuting operators.  This can be thought of as a `ghost' of the
previous geometric symmetry.  These higher symmetries generate an algebra with
{\it nonlinear} commutation relations, whose representations are built in a way
analogous to the highest weight representations of Lie algebras. In $2$
dimensions the symmetry algebra is very small, being just $sl(2,\mathbb{C})$ in
our case (the 2-dimensional Euclidean algebra in the f\/lat case), so the
representations are very simple.  In higher dimensions, the usual highest
weight representations apply for the symmetry algebra of the Laplace--Beltrami
operator.  In this case there is no longer a single chain of eigenfunctions, so
we can expect similar behaviour for the algebra of higher symmetries.

\pdfbookmark[1]{Appendix. Eigenfunctions of the Laplace-Beltrami operator}{appendix}
\section*{Appendix. Eigenfunctions of the Laplace--Beltrami operator}

We use the {\it highest weight representations} of the symmetry algebra
$sl(2,\mathbb{C})$ to construct eigenfunctions of $L_b$.  Since $\bf H$ and
$L_b$ commute, they share eigenfunctions and for $\bf H$ these are built by the
highest weight construction.  Furthermore, starting with {\it any}
eigenfunction of $L_b$, we may use the symmetry algebra to construct further
eigenfunctions (with the same eigenvalue).  Since this eigenspace is {\it
invariant} under the action of $sl(2,\mathbb{C})$ (by construction), it can be
decomposed into irreducible components, which are just weight spaces. Therefore
{\it all} eigenfunctions of $L_b$ can be written as linear combinations of
those we construct below.

A {\it highest weight vector} $\psi_1^n$, of weight $2 n$, satisf\/ies
\begin{gather*}
{\bf E}\, \psi_1^n = 0, \qquad {\bf H}\, \psi_1^n = 2 n \psi_1^n,
\end{gather*}
which constitute a pair of partial dif\/ferential equations for the
eigenfunction.  These are compatible on the zeros of the dif\/ferential operator
$\bf E$, since
\begin{gather*}
{\bf H}\, {\bf E}\, \psi_1^n - {\bf E}\, {\bf H}\, \psi_1^n =
          2 {\bf E}\, \psi_1^n = 0.
\end{gather*}
The specif\/ic form of $\psi_1^n$ depends upon $n$ and upon the choice of
representation for $sl(2,\mathbb{C})$.  However, the general structure of the
representation is {\it independent} of this specif\/ic form, being a consequence
only of the commutation relations (\ref{sl2comm}) (see \cite{72-3}).

Def\/ining $\psi_r^n = {\bf F}^{r-1} \psi_1^n$, the commutation relations imply: %
\begin{gather}
\label{module}  %
{\bf H}\, \psi_r^n = 2 (n+1-r) \psi_r^n,\qquad
{\bf E}\, \psi_r^n = (r-1) (2 n+2-r) \, \psi_{r-1}^n.  %
\end{gather} %
Our def\/inition of Laplace--Beltrami operator $L_b$ as Casimir operator
(\ref{cas}) implies that
\begin{gather*}
L_b\, \psi_r^n = \frac{1}{4} \, n(n+1) \psi_r^n,\qquad \mbox{for all}\ \ r, \ n.
\end{gather*}

We can also construct a 3-point recursion relation between these
eigenfunctions, but this {\it is} explicitly dependent upon the form of the
operators. Let
\begin{gather*}
{\bf H} = h_1 \, \pa_{z_1} + h_2\, \pa_{z_2},\qquad
          {\bf E} = e_1 \, \pa_{z_1} + e_2\, \pa_{z_2},\qquad
             {\bf F} = f_1 \, \pa_{z_1} + f_2\, \pa_{z_2},
\end{gather*}
where $h_i$, etc are functions of the coordinates $z_i$.  We have (with $\mu_r$
and $a_r$ def\/ined by (\ref{module}))  %
\begin{gather*}
\left. \begin{array}{l}
{\bf H}\, \psi_r^n = \mu_r \psi_r^n,
\\[1ex]
{\bf E}\, \psi_r^n = a_r \psi_{r-1}^n
\end{array}  \right\} \quad\Rightarrow\quad
\left(\begin{matrix}
\pa_{z_1} \psi_r^n
\\[1ex]
\pa_{z_2} \psi_r^n
\end{matrix}  \right) =  \frac{1}{h_1 e_2- e_1 h_2}
\left(\begin{matrix}
e_2 & -h_2
\\[1ex]
-e_1 & h_1
\end{matrix}  \right)
\left(\begin{matrix}
\mu_r \psi_r^n
\\[1ex]
a_r \psi_{r-1}^n
\end{matrix}  \right).     %
\end{gather*}
The relation $\psi_{r+1}^n = {\bf F}\, \psi_r^n$ then implies that  %
\begin{gather*}
\psi_{r+1}^n = \frac{f_1 e_2-e_1 f_2}{h_1 e_2- e_1 h_2} \mu_r \psi_r^n +
\frac{h_1 f_2-f_1 h_2}{h_1 e_2- e_1 h_2} a_r \psi_{r-1}^n.
\end{gather*} %
When this is singular the representation reduces to 1 dimension.

In the case of our vector f\/ield representation (\ref{sl2-1}), this takes
explicit form
\begin{gather*}
\psi_{r+1}^n = 2(n-r+1) \sqrt{\frac{y}{x}}\, \psi_r^n
         +(r-1)(2n+2-r)\left(\frac{y-1}{x}\right) \psi_{r-1}^n,
\end{gather*}
with
\begin{gather*}
\psi_1^n = \sqrt{x^n},\qquad \psi_2^n = 2 n \sqrt{x^{n-1}y}.
\end{gather*}

When $n$ is an integer, these representations are of f\/inite dimension $2 n+1$
and irreducible, but inf\/inite dimensional otherwise.  For instance, when $n=2$,
we have
\begin{gather*}
\psi_1^2=x,\qquad \psi_2^2=4 \sqrt{xy},\qquad \psi_3^2=4(3y-1),\\
\psi_4^2=24(y-1)\sqrt{\frac{y}{x}},\qquad \psi_5^2=\frac{24(y-1)^2}{x}.
\end{gather*}

\pdfbookmark[1]{References}{ref}
\LastPageEnding


\begin{thebibliography}{99}

\footnotesize\itemsep=0pt

\bibitem{83-9}
Boyer~C.P., Kalnins E.G., Winternitz P.,
 Completely integrable relativistic {Hamiltonian} systems and
  separation of variables in Hermitian hyperbolic spaces,
 {\it J. Math. Phys.} \textbf{24} (1983), 2022--2034.

\bibitem{01-5}
Daskaloyannis~C.,
Quadratic Poisson algebras of two-dimensional classical
superintegrable systems and quadratic associative algebras of quantum
superintegrable systems, {\it J. Math. Phys.} \textbf{42} (2001), 1100--1119,
\href{http://arxiv.org/abs/math-ph/0003017}{math-ph/0003017}.

\bibitem{f05-1}
Fordy A.P.,  Symmetries, ladder operators and quantum integrable systems,
{\it Glasg. Math. J.} \textbf{47} (2005), 65--75.

\bibitem{f06-1}
Fordy A.P.,  Darboux related quantum integrable systems on a constant curvature
surface,  {\it J. Geom. Phys.} \textbf{56} (2006), 1709--1727.

\bibitem{74-7}
Gilmore~R.,
Lie groups, Lie algebras and some of their applications,
Wiley, New York, 1974.

\bibitem{01-1}
Harnad~J., Vinet~L., Yermolayeva~O., Zhedanov~A.,
Two-dimensional Krall--Shef\/fer polynomials and integ\-rable systems,
{\it J. Phys. A: Math. Gen.} \textbf{34} (2001), 10619--10625.

\bibitem{72-3}
Humphreys J.E.,
Introduction to Lie algebras and representation theory,
 Springer-Verlag, Berlin, 1972.

\bibitem{51-1}
Infeld~L., Hull~T., The factorization method,
{\it Rev. Modern Phys.} \textbf{23} (1951), 21--68.

\bibitem{01-7}
Kalnins E.G., Kress J.M., Pogosyan G.S., Miller W.Jr.,
 Completeness of superintegrability in two-dimensional
  constant-curvature spaces,
 {\it J. Phys. A: Math. Gen.} \textbf{34} (2001), 4705--4720, \href{http://arxiv.org/abs/math-ph/0102006}{math-ph/0102006}.

\bibitem{99-11}
Kalnins E.G., Miller W.Jr., Hakobyan Ye.M., Pogosyan G.S.,
 Superintegrability on the two-dimensional hyperboloid.~II,
 {\it J. Math. Phys.} \textbf{40} (1999), 2291--2306,
 \href{http://arxiv.org/abs/quant-ph/9907037}{\mbox{quant-ph/9907037}}.

\bibitem{06-1}
Kalnins E.G., Miller  W.Jr., Pogosyan G.S.,
 Exact and quasiexact solvability of second-order superin\-teg\-rable
  quantum systems. I.~Euclidean space preliminaries,
 {\it J. Math. Phys.} \textbf{47} (2006), 033502, 30 pages, \href{http://arxiv.org/abs/math-ph/0412035}{\mbox{math-ph/0412035}}.

\bibitem{67-2}
Krall H.L., Shef\/fer I.M.,
 Orthogonal polynomials in two variables,
 {\it Ann. Mat. Pura Appl. (4)} \textbf{76} (1967), 325--376.

\bibitem{96-5}
Kuznetsov V.B.,
 Hidden symmetry of the quantum Calogero--Moser system,
 {\it Phys. Lett. A} \textbf{218} (1996), 212--222, \href{http://arxiv.org/abs/solv-int/9509001}{solv-int/9509001}.

\bibitem{01-8}
Tempesta~P., Turbiner A.V., Winternitz~P.,
 Exact solvability of superintegrable systems,
 {\it J. Math. Phys.} \textbf{42} (2001), 4248--4257, \href{http://arxiv.org/abs/hep-th/0011209}{hep-th/0011209}.

\bibitem{03-4}
Vinet~L., Zhedanov~A.,
 Two-dimensional Krall--Shef\/fer polynomials and quantum systems on
  spaces of constant curvature,
 {\it Lett. Math. Phys.} \textbf{65} (2003), 83--94.

\end{thebibliography}
\end{document}